\documentclass[aps,prb,twocolumn,showpacs,floatfix,superscriptaddress]{revtex4}
\usepackage{graphicx,graphics}
\usepackage{natbib}
\usepackage{lipsum}
\usepackage{dcolumn}
\usepackage{amsmath,amssymb,amsfonts}
\usepackage{latexsym,verbatim}
\usepackage{bm}
\usepackage{color}
\usepackage{ulem}
\usepackage{siunitx}
\usepackage[breaklinks=true,colorlinks,citecolor=blue,linkcolor=blue,urlcolor=blue]{hyperref}

\begin{document}
\title{Generation of multiple plasmons in strontium niobates mediated by local field effects}
\author{Tao Zhu}
\affiliation{Department of Physics, National University of Singapore, Singapore 117542, Singapore}
\affiliation{Singapore Synchrotron Light Source, National University of Singapore, Singapore 117603, Singapore}
\author{Paolo E. Trevisanutto}
\affiliation{Department of Physics, National University of Singapore, Singapore 117542, Singapore}
\affiliation{Singapore Synchrotron Light Source, National University of Singapore, Singapore 117603, Singapore}
\affiliation{Center for Advanced 2D Materials, National University of Singapore, 117542, Singapore}
\author{Teguh Citra Asmara}
\affiliation{Singapore Synchrotron Light Source, National University of Singapore, Singapore 117603, Singapore}
\affiliation{NUSNNI-NanoCore, National University of Singapore, Singapore 117411, Singapore}
\author{Lei Xu}
\affiliation{Department of Physics, National University of Singapore, Singapore 117542, Singapore}
\author{Yuan Ping Feng}
\affiliation{Department of Physics, National University of Singapore, Singapore 117542, Singapore}
\affiliation{Center for Advanced 2D Materials, National University of Singapore, 117542, Singapore}
\author{Andrivo Rusydi}
\email{phyandri@nus.edu.sg}
\affiliation{Department of Physics, National University of Singapore, Singapore 117542, Singapore}
\affiliation{Singapore Synchrotron Light Source, National University of Singapore, Singapore 117603, Singapore}
\affiliation{Center for Advanced 2D Materials, National University of Singapore, 117542, Singapore}
\affiliation{NUSNNI-NanoCore, National University of Singapore, Singapore 117411, Singapore}
\affiliation{NUS Graduate School for Integrative Sciences and Engineering, National University of Singapore, Singapore 117456, Singapore}

\begin{abstract}
	Recently, an anomalous generation of multiple plasmons with large spectral weight transfer in the visible to ultraviolet range (energies below the band gap) has been experimentally observed in the insulating-like phase of oxygen-rich strontium niobium oxides (SrNbO$_{3+\delta}$). Here, we investigate the ground state and dielectric properties of SrNbO$_{3+\delta}$ as a function of $\delta$ by means of extensive first principle calculations. We find that in the random phase approximation by taking into account the local field effects (LFEs), our calculations are able to reproduce both the unconventional multiple generations of plasmons and spectral weight transfers, consistent with experimental data. Interestingly, these unconventional plasmons can be tuned by oxygen stoichiometry as well as microscopic superstructure. This unusual predominance of LFEs in this class of materials is ascribed to the strong electronic inhomogeneity and high polarizability and paves a new path to induce multiple plasmons in the untapped visible to ultraviolet ranges of insulating-like oxides.
\end{abstract}
\maketitle

\section{Introduction}
The Sr$_{1-x}$Nb$_{1-y}$O$_{3+\delta}$ oxides have been known for their rich properties, which strongly depend on their oxygen stoichiometry \cite{ridgley, isawa, weber, sakai, kobayashi, lichtenberg,kuntscher1, kuntscher2, chen}. Upon increasing their oxygen content, the oxides can be transformed from three-dimensional (3D) metal SrNbO$_3$, to quasi-one-dimensional metal SrNbO$_{3.4}$, and finally to ferroelectric insulator SrNbO$_{3.5}$. Recently, these oxides have been reported to exhibit several exotic phenomena based on their anomalous plasmon behaviors. The metallic perovskite SrNbO$_3$, for instance, has been found as a potentially good visible-light photocatalyst for water splitting applications \cite{xu}, and this photoactivity has been reported to be driven by plasmon-based resonances \cite{wan} instead of the more usual interband excitations. More interestingly, under oxygen enrichment, the SrNbO$_{3+\delta}$ has been reported to exhibit an anomalous multiple bulk plasmons generation \cite{asmara}.

These plasmons have shown several unconventional features \cite{asmara}. They have been detected, intriguingly, in the insulating-like samples with room temperature resistivity up to 6 $\Omega$ cm. As a function of $\delta$, they have became weaker and eventually emerged into a single plasmon in samples with higher conductivity. Moreover, due to the high resistivity, the real part of the complex dielectric function, $\varepsilon_1$, has no negative value below the plasmon energies. These indicate that these plasmons are different from the conventional bulk plasmons as the later typically shows a zero cross of the real dielectric function. Finally, the losses of these plasmons are fairly low, i.e. several times lower than that in gold, which provides a new direction for the plasmonic research in the visible-ultraviolet ranges.

The origin of these unconventional plasmons remains an open question and a proper theoretical description is still lacking. Previously, Asmara \textit{et al}. \cite{asmara} suggested a possible explanation by using a phenomenological semi-classical model with adjustable parameters. On other hand, calculations based on dynamical mean field theory (DMFT) framework \cite{dmft1,dmft2}, have displayed that in strongly correlated systems, qualitatively  to some extent, similar spectral weight transfers and generation of multiple anomalous plasmons can occur \cite{vanloon}. However, these theoretical results were not wholly consistent with the experimental findings \cite{asmara}, in particular regarding both the predicted energy ratio of the plasmon peak energies and how oxygen doping may affect the formation of plasmons. More importantly, the previously calculated absorption spectra were obtained by neglecting the microscopic fluctuations that may give rise to local field effects (LFEs) \cite{adler,wiser,lfe1,lfe2,lfe3,lfe4,lfe5}. This was despite the importance of LFEs in modifying the shape and intensity of loss function spectra in various materials, particularly at high energies where semi-core localized electrons involved\cite{vast}.

In this article, we present a systematic study of the mechanisms and properties of plasmons in strontium niobates by using \textit{ab initio} calculations. Within the linear response time-dependent density functional theory (LR-TDDFT) framework \cite{gajdos}, we have calculated the complex dielectric and the loss functions of SrNbO$_{3+\delta}$ as a function of oxygen doping and its microscopic superstructures. In the random phase approximation (RPA) \cite{rpa} and taking into account LFEs , we show that the fundamental properties of this multiple plasmons formation and their evolutions are well described and in good agreement with experimental findings \cite{asmara}. For SrNbO$_3$, the experimental plasmon is reproduced and it is described as an interplay of interband and intraband transitions. LFEs do not play any role. In contrast, for oxygen-rich SrNbO$_{3+\delta}$, the LFEs enhancement yields drastic spectral weight transfers and multiple unconventional plasmons are rising. Our analysis shows that the LFEs stem from the localization of Nb 4$d$ electrons in cooperation with the high polarizability of the oxygen-rich SrNbO$_{3+\delta}$ systems.

The article is divided into five sections. In Section II, we introduce the theoretical methods and parameters used in the first principle calculations. The structure of two typical SrNbO$_{3+\delta}$ and the DFT based RPA method with its inclusion of LFEs are discussed. The calculated results for electronic structure and dielectric properties of SrNbO$_3$ and two configurations of SrNbO$_{3.4}$ are shown in Sec. III. In Sec. IV, we propose a theoretical explanation for the experimental detection of these multiple plasmons and we analyze the importance of the extra-oxygen-plane (EOP) in determining the creation of these unconventional multiple plasmons. The last section summarizes and concludes the main achievements of this work.

\section{Methods}

\begin{figure}
	\includegraphics[width=8.6 cm]{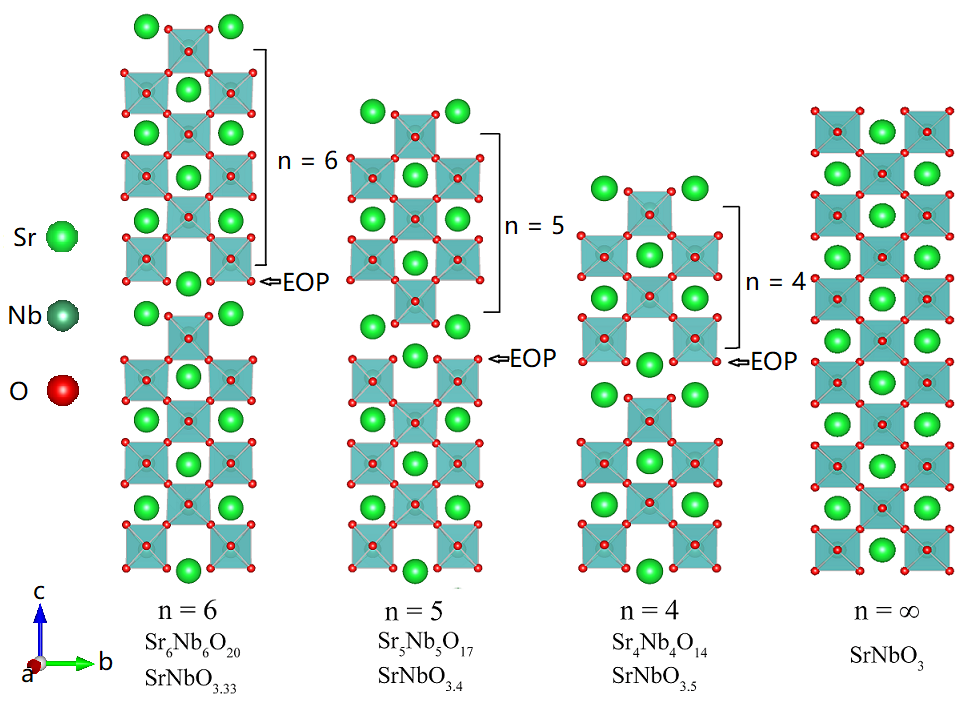}
	\caption{Sketch of crystal structures of the perovskite-related layered series Sr$_{n}$Nb$_n$O$_{3n+2}$ for n = 6, 5, 4 and $\infty$. The extra oxygen planes (EOP) are shown in between two neighboring layers.}
	\label{fig1}
\end{figure}

As representative cases, we study the effects of LFEs on SrNbO$_{3+\delta}$ as a function of oxygen doping, i.e. SrNbO$_3$ and SrNbO$_{3.4}$, and microscopic superstructure of alternating layers from SrNbO$_{3.33}$ and SrNbO$_{3.5}$. Their complex dielectric functions, $\varepsilon=\varepsilon_1+i\varepsilon_2$, and loss functions, -Im$[\varepsilon^{-1}]$, are calculated with and without LFEs, and then compared with experimental data \cite{asmara}. In terms of structure, ideal structure of SrNbO$_3$ is created by a 3D network of corner-sharing NbO$_6$ octahedra with Sr atoms filling in the center of the network. With oxygen enrichment, however, the structure of SrNbO$_{3.4}$ becomes homologous to layered perovskite series of Sr$_{n}$Nb$_n$O$_{3n+2}$ family as shown in Fig. \ref{fig1}. These series can be derived from the 3D network of the SrNbO$_{3}$ perovskite structure by separating the NbO$_{6}$ octahedra with EOP parallel to the [101] perovskite planes every \textit{n} unit cells, creating a layered structure with stacks of slabs along their c-axis \cite{lichtenberg}. This makes the structure of SrNbO$_{3.4}$ strongly anisotropic which creates strong charge inhomogeneity and high polarizability along one particular crystal orientation \cite{kuntscher1,kuntscher2}.

We began with DFT calculations of the ground state property of SrNbO$_{3+\delta}$ by using the Vienna \textit{ab initio} simulation package (VASP) \cite{Vasp1,Vasp2}. The projector-augmented wave (PAW) method \cite{paw} and the Perdew-Burke-Ernzerhof (PBE) exchange-correlation functional within GGA \cite{gga} were employed. Plane-wave basis set with \SI{500}{\electronvolt} energy cut-off was used to expand the Kohn-Sham wave functions \cite{ks}. The Sr $4s4p5s$, Nb $4p5s4d$, and O $2s2p$ orbitals were treated as valence states. The irreducible Brillouin zone (BZ) of cubic SrNbO$_3$ was sampled by a $13\times13\times13$ $\Gamma$ centered Monkhorst-Pack \cite{mp} grid and a $9\times7\times1$ $\Gamma$ centered Monkhorst-Pack grid has been employed for the supercell of layered SrNbO$_{3+\delta}$. The Gaussian smearing with the width of \SI{0.05}{\electronvolt} was adopted to treat the partial occupancies. The lattice constant of the fully relaxed structure of SrNbO$_3$ is $a = b = c = \SI{4.07}{\SIUnitSymbolAngstrom}$ and the optimized lattice parameter of the supercell of SrNbO$_{3.4}$ is $a = \SI{4.03}{\SIUnitSymbolAngstrom}$, $b = \SI{5.75}{\SIUnitSymbolAngstrom}$, and $c = \SI{33.45}{\SIUnitSymbolAngstrom}$. These values are in good agreement with those reported in previous studies. \cite{kuntscher2, chen}.

The dielectric matrix is calculated on top of the ground state calculations. We further follow with a simple RPA approach in which the Fourier transform of the symmetric microscopic dielectric matrix is given by
\begin{equation}
\varepsilon_{{\bf G},{\bf G}'}({\bf q},\omega)=\delta_{{\bf G},{\bf G}'}-v({\bf q}+{\bf G})\chi_{{\bf G},{\bf G}'}^{0}({\bf q},\omega)
\label{dielectric function}
\end{equation}
where $\bf{G}$ and $\bf{G}^\prime$ are reciprocal lattice vectors, $\bf{q}$ is the Bloch vector in the first Brillouin zone and $v$ is the bare Coulomb interaction. The independent particle irreducible polarizability matrix, $\chi_{\bf{G},\bf{G}^\prime}^{0}(\bf{q},\omega)$, is summed over transitions between occupied and empty states and can be constructed from Kohn-Sham wavefunctions $\psi$ and eigenvalues $\epsilon$ \cite{Onida, gajdos}.

The macroscopic dielectric function (which is comparable to experiment) is determined at the limit of $\bf{q}\to0$ as
\begin{equation}
\varepsilon_M(\omega)=\lim\limits_{\bf{q}\to 0}\dfrac{1}{\varepsilon^{-1}_{\bf{G}=0,\bf{G}^\prime=0}(\bf{q},\omega)}.
\label{macro}
\end{equation}
The energy loss function are then given by -Im$[{\varepsilon^{-1}_M}]$ for vanishingly small wave vector \cite{Onida}.

The off-diagonal elements of the dielectric function in the matrix inversion of Eq. (\ref{macro}), $\varepsilon^{-1}_{\bf{G},\bf{G}^\prime}(\bf{q},\omega)$, are responsible for the LFEs and they become important in inhomogeneous systems where localization of atomic orbitals play a major role (like in \textit{d} and \textit{f} transition metals). Alternatively, the LFEs can be included in $\varepsilon_M$ explicitly by constructing a modified response function \cite{Onida}
\begin{equation}
\varepsilon_M(\omega)=1-\lim\limits_{\bf{q}\to 0}v_0(\bf{q})\bar{\chi}_{\bf{G}=0,\bf{G}^\prime=0}(\bf{q},\omega),
\label{dyson}
\end{equation}
where we have introduced $\bar{\chi}$ that satisfies a Dyson-like equation of $\bar{\chi}=\chi^0+\chi^0\bar{v}\bar{\chi}$ and $\bar{v}$ is the bare Coulomb interaction (excluding the long range term in $v$). In the Dyson-like equation for $\bar{\chi}$, the second term with $\bar{v}$ plays the role of LFEs \cite{Onida,sottile}. Neglecting this term in the expression of $\bar{\chi}$ yields the so-called independent particle random phase approximation (IP-RPA):
\begin{eqnarray}
\nonumber\varepsilon_M(\omega)&=&1-\lim\limits_{\bf{q}\to0}v_0(\bf{q})\chi^0_{\bf{G}=0,\bf{G}^\prime=0}(\bf{q},\omega)\\
&=&\lim\limits_{\bf{q}\to0}\varepsilon_{\bf{G}=0,\bf{G}^\prime=0}(\bf{q},\omega)
\label{nlf}
\end{eqnarray}
where the off-diagonal elements of the matrix inversion of Eq. (\ref{macro}) are neglected. In addition, we have also included the intraband transitions in the RPA polarizability by using the DP code \cite{dp} to quantify the effects of the Drude tail.

\section{Results}\label{Results}

\subsection{SrNbO$_3$}
\begin{figure}
	\includegraphics[width=8.6 cm]{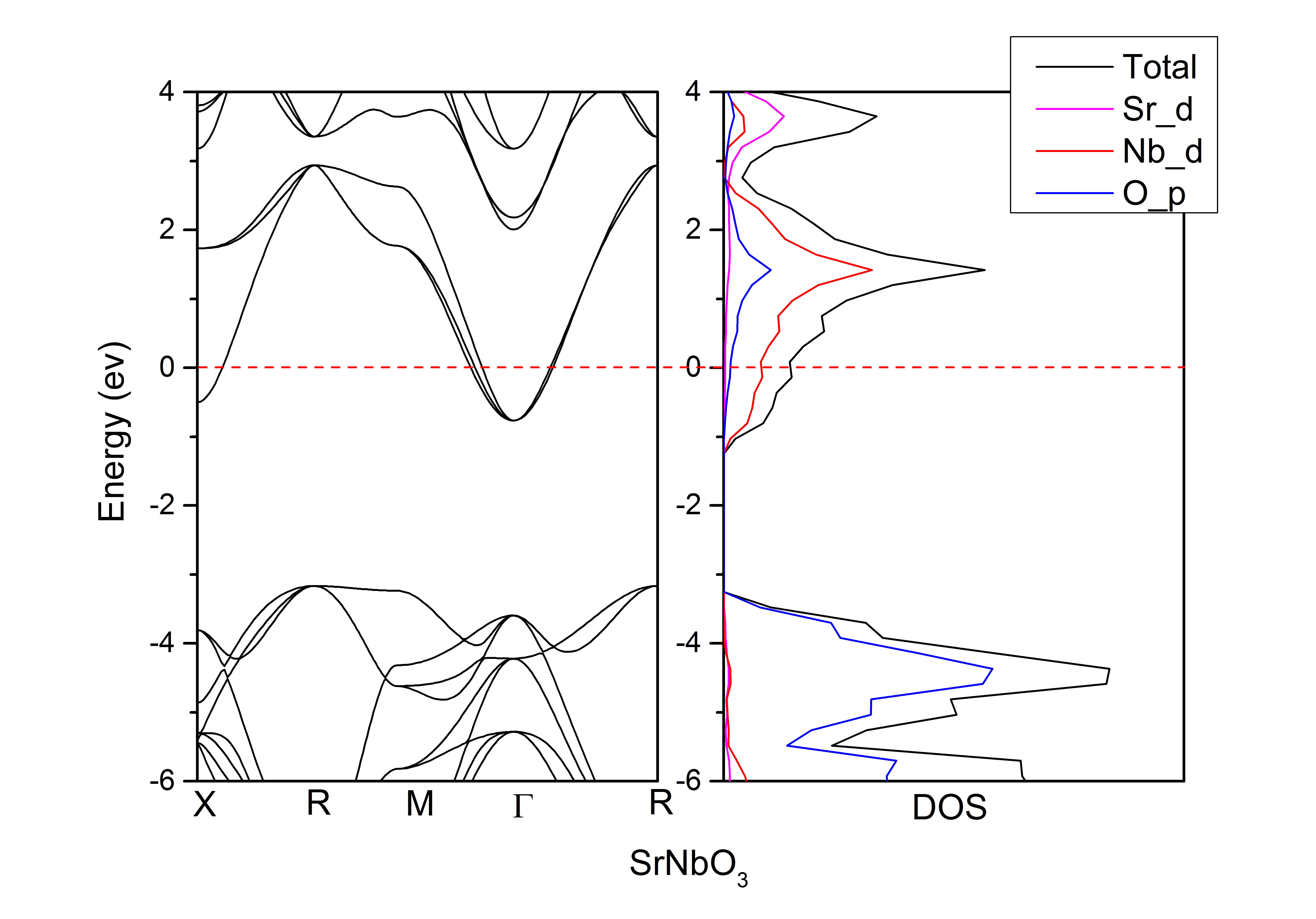}
	\caption{Band structure and density of states of SrNbO$_3$. Red dashed line represents the Fermi level.}
	\label{fig2}
\end{figure}
The calculated band structure and density of states (DOS) of SrNbO$_3$ are shown in Fig. \ref{fig2}. The SrNbO$_3$ is a metallic material as the Fermi energy ($E_F$) crosses the conduction band. The Nb $4d$ orbitals which split into the triply degenerated $t_{2g}$ and the doubly degenerated $e_g$ in the octahedral field are responsible for its metallicity. Thus, little distortion of the NbO$_6$ octahedra in SrNbO$_3$ has been reported\cite{chen}. Interestingly, an apparent large indirect band gap ($\sim$ \SI{2.3}{\electronvolt}) can be seen just below the $E_F$. The conduction band minimum (CBM) lies at the $\Gamma$ point while the valence band maximum (VBM) located at the R point. The existence of this large band gap below the $E_F$ suggests that the photocatalytic activity of SrNbO$_3$ may arise from the hot electrons generated from the decay of the plasmon resonance instead of the interband transitions \cite{wan}. From the site-projected DOS (PDOS), it can be seen that the oxygen $p$ orbital contributed most of the DOS just below the Fermi energy and the chemical bonding exists between electrons from O 2$p$ orbitals and Nb 4$d$ orbitals.
\begin{figure}
	\includegraphics[width=8.6 cm]{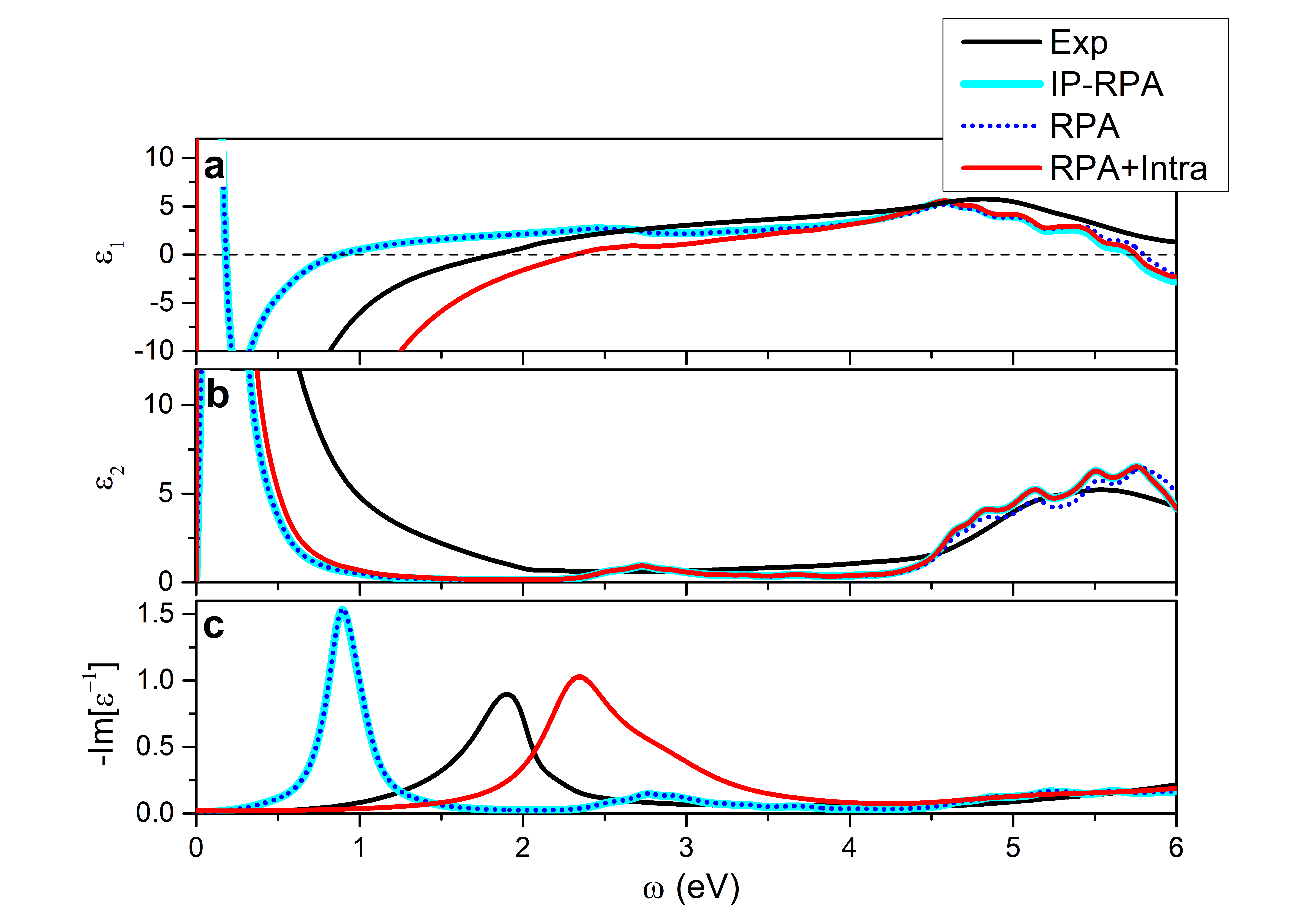}
	\caption{Calculated macroscopic complex dielectric and loss functions of SrNbO$_3$. The calculated (a) real part, $\varepsilon_1$, and (b) imaginary part, $\varepsilon_2$, of complex dielectric function as well as (c) loss function, -Im$[\varepsilon^{-1}]$, of SrNbO$_3$ are compared with those of experiment (Exp) \cite{asmara}. RPA and IP-RPA refer to the calculation with and without LFEs, respectively (they almost overlapped). RPA + Intra refers to RPA calculation which includes intraband transitions. The horizontal dashed line indicates the zero of $\varepsilon_1$ to pinpoint the conventional plasmon peak energy.}
	\label{fig3}
\end{figure}

Figure \ref{fig3} shows the calculated complex dielectric and loss functions of SrNbO$_{3}$ compared with experimental data from the Sr$_{1-x}$NbO$_{3+\delta}$ film deposited under low oxygen pressure (labeled as lp-SNO in Ref. \citenum{asmara}). As seen, in the experimental results, metallic SrNbO$_3$ has a conventional single plasmon resonance at $\sim$ \SI{1.9}{\electronvolt}, which coincides with the zero of $\varepsilon_1$. This apparent plasmon peak located at $\sim$ \SI{1.9}{\electronvolt} has similar energy with the absorption edge in photocatalysis experiment which implies its plasmonic origin \cite{wan}. The IP-RPA calculations are sufficient to reproduce this main feature. Since IP-RPA calculations regard the electrons as independent, non-interacting particles, this indicates that the partially-filled 4\textit{d} orbitals do not induce strong correlations in SrNbO$_3$, making it a weakly-correlated material. The inclusion of LFEs also does not change the calculated spectra, indicating that LFEs do not play a role here and SrNbO$_3$ is electrically homogeneous. The underestimation of the conventional plasmon resonance energy in the IP-RPA calculation can be corrected by adding intraband transitions of Nb 4\textit{d} orbitals, in addition to the interband transitions from O 2\textit{p} to Nb 4\textit{d}. Thus, the plasmon at $\sim$ \SI{1.9}{\electronvolt} in SrNbO$_3$ originates from the cumulative effects of interband and intraband transitions.
\begin{figure}
	\includegraphics[width=8.6 cm]{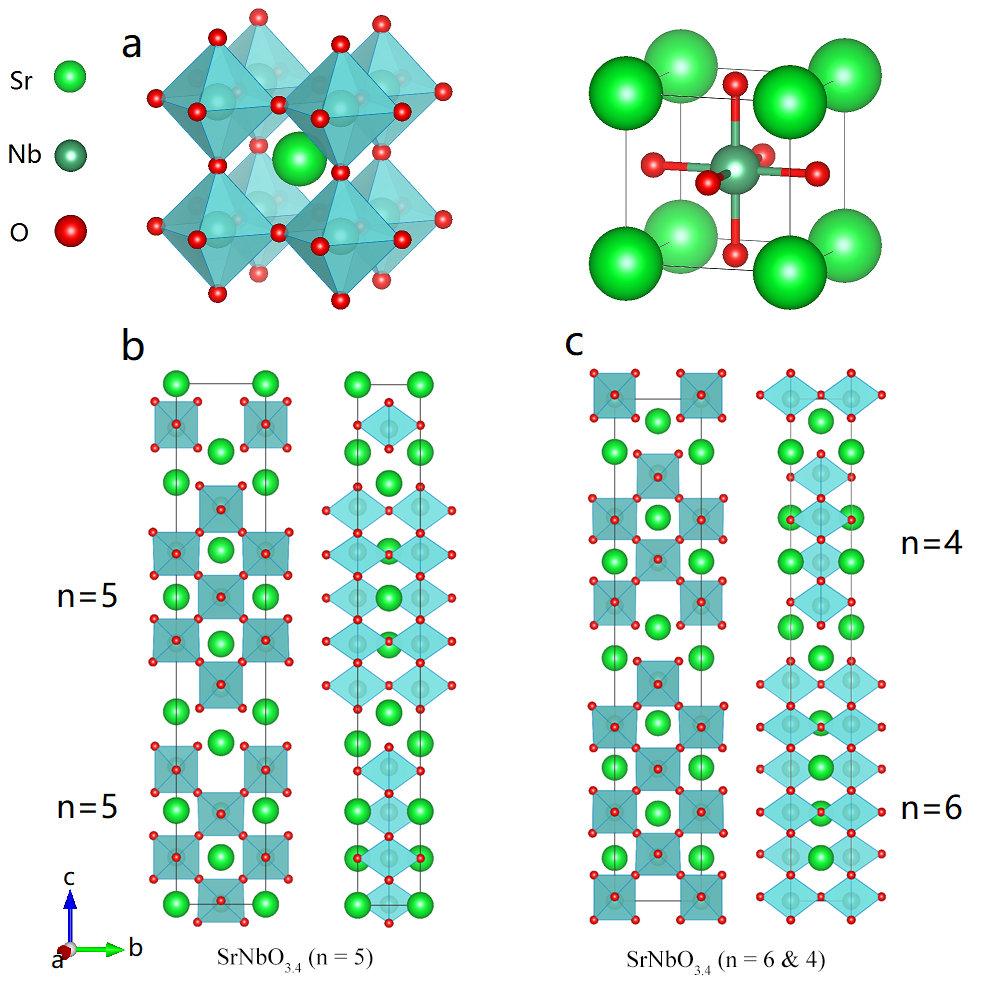}
	\caption{Crystal structure of SrNbO$_3$ and two configurations of SrNbO$_{3.4}$. (a) SrNbO$_3$, (b) SrNbO$_{3.4}$ (n = 5), and (c) SrNbO$_{3.4}$ (n = 6, 4, ...) projected along the a- and b-axis. In each layer, the connection of NbO$_6$ octahedra behaves zig-zag-like along the b-axis and chain-like along the a-axis. Every nearest-neighbor layer has a height difference of a half of the NbO$_6$ octahedron body diagonal along the a-axis, while every next-nearest-neighbor layer has the same height.}
	\label{fig4}
\end{figure}
\subsection{SrNbO$_{3.4}$ (n = 5)}
\begin{figure}
 	\includegraphics[width=8.6 cm,height=8cm]{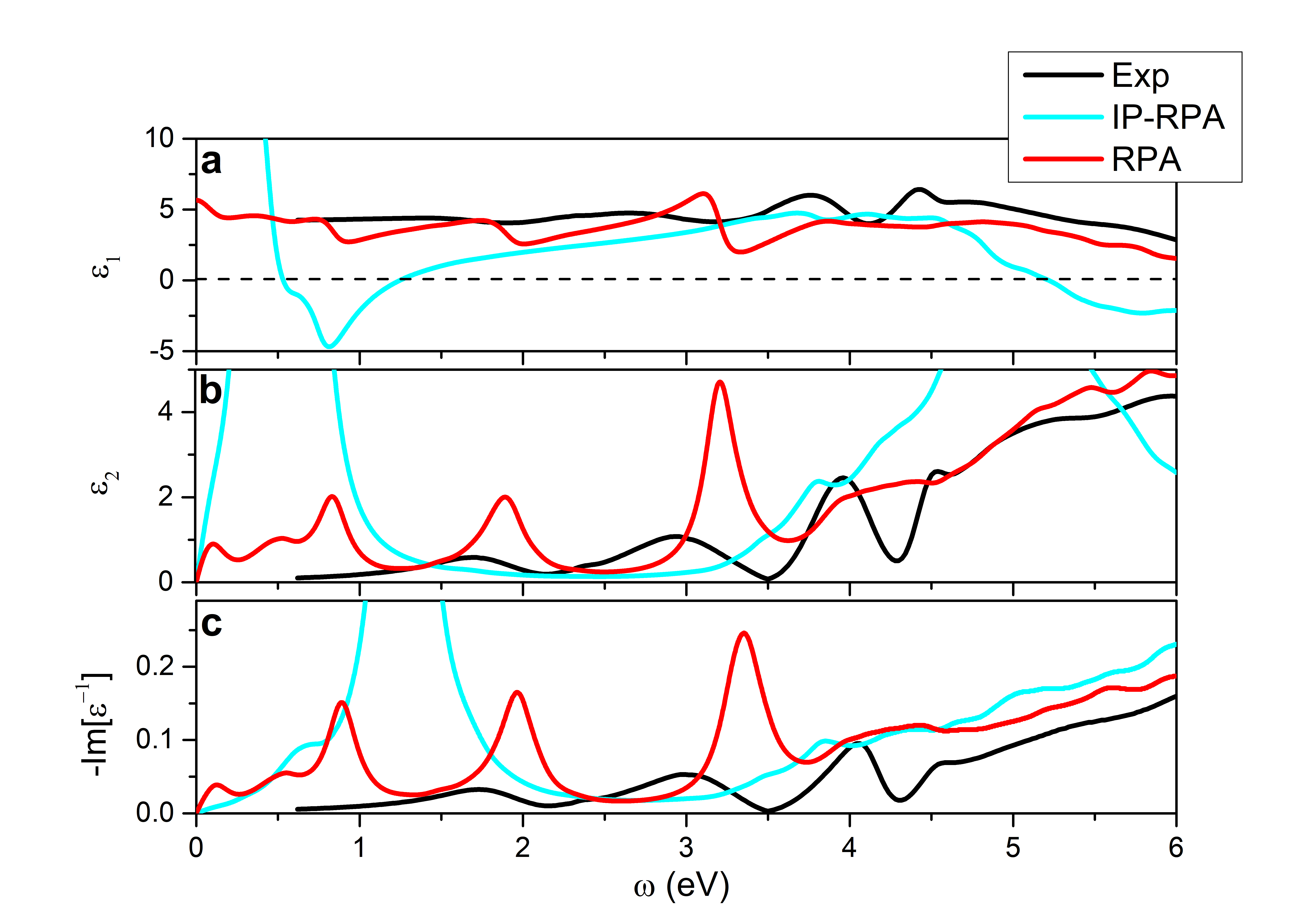}
 	\caption{Calculated macroscopic complex dielectric and loss functions of SrNbO$_{3.4}$ along the c-axis. The calculated (a) real part, $\varepsilon_1$, and (b) imaginary part, $\varepsilon_2$, of complex dielectric function as well as (c) loss function, -Im$[\varepsilon^{-1}]$, of SrNbO$_{3.4}$ are compared with those of experiment (Exp) \cite{asmara}. RPA and IP-RPA refer to the calculation with and without LFEs, respectively. The horizontal dashed line indicates the zero of $\varepsilon_1$ to pinpoint the conventional plasmon peak energy.}
 	\label{fig5}
 \end{figure}

Because of the experimental sample shows a layered structure in which every few NbO$_6$ octahedra are separated by the EOPs. We calculated the dielectric function and loss function of SrNbO$_{3.4}$ along the c-axis as shown in Fig. \ref{fig5}. We can see that the situations for SrNbO$_{3.4}$, on the other hand, are totally different. Because SrNbO$_{3.4}$ is nominally not fully oxidized, IP-RPA again predicts the existence of a metallic conventional plasmon peak. However, the experimental data have shown that the Drude tail was absent in the $\varepsilon_2$ spectrum and the $\varepsilon_1$ spectrum remained positive at all energy ranges, indicating it has an insulating-$like$ behavior. Furthermore, unlike for SrNbO$_{3}$, the conventional single plasmon peak is not present; instead, unconventional multiple plasmon peaks have emerged at $\sim$ \SI{1.7}{\electronvolt}, $\sim$ \SI{3.0}{\electronvolt}, and $\sim$ \SI{4.0}{\electronvolt} accompanied by spectral weight transfers.

These fundamental differences between SrNbO$_{3.4}$ and SrNbO$_{3}$ suggest that the mechanisms behind these plasmons are very different from conventional plasmons in metals. It has been shown in the previous theoretical study \cite{vanloon} that vertex corrections might indeed induce spectral weight transfers and created multiple plasmon peaks in strongly-correlated systems. However, as shown in Fig. \ref{fig3}, the partially filled Nb 4$d$ orbitals do not induce strong electron correlations in SrNbO$_3$. More importantly, the predicted energy and ratio of plasmon peaks are different from that of the observed unconventional plasmons in SrNbO$_{3.4}$ \cite{asmara}, suggesting that other mechanisms should be involved. On the other hand, it has been recently shown that in the absorption spectrum of correlated materials, such as VO$_{2}$, LFEs have strong contribution into the optical spectra due to the localized nature of electron-hole pairs \cite{gatti}. In the c-axis, SrNbO$_{3.4}$ has a layered structure generated by the EOP (Fig. \ref{fig4}) which confine the Nb 4$d$ electrons stay within each slab and further create a strong electronic inhomogeneity and high polarizability along the c-axis. Therefore the strong modification of the spectrum from LFEs may be expected.

Our main theoretical result, as shown in  Fig. \ref{fig5}, is that LFEs are surprisingly dominant in the SrNbO$_{3.4}$ complex dielectric and loss function spectra. When the LFEs are introduced into the RPA calculations, the conventional plasmon at $\sim$ \SI{1.2}{\electronvolt} is suppressed and new unconventional multiple plasmons at $\sim$ \SI{1.9}{\electronvolt} and $\sim$ \SI{2.9}{\electronvolt} are generated, instead. The calculated $\varepsilon_1$ also becomes positive at almost all energy ranges which indicates its insulating feature. All of these are consistent with the experimental data \cite{asmara}.

\subsection{SrNbO$_{3.4} (n = 4, 6, ...)$}
From Eq. (\ref{dyson}), it can be understood that, in addition to the electrical inhomogeneity, strong LFEs can also originate from a large Kohn-Sham polarizability $\chi^0$, for example in strongly polarized systems with very localized orbitals \cite{gatti,lfe5}. To further investigate the effects of high polarizability and also to account for possible non-uniformities in the experimental SrNbO$_{3.4}$ sample \cite{asmara}, we performed calculations on an intercalated SrNbO$_{3.4}$ structure as described below. Within the layered Sr$_{n}$Nb$_n$O$_{3n+2}$ series, SrNbO$_{3.4}$ is formed when n = 5, which means that adjacent EOP are separated by a layer of 5 NbO$_6$ octahedra thick along the $c$-axis. In the intercalated structure, the stacking sequence is tweaked so that the layers are intercalated in 4 and 6 NbO$_6$ octahedra thicknesses instead, i.e. n = 4, 6, 4, 6, etc. As shown in Fig. \ref{fig4}, this alternate configuration can be created by combining layers from metallic SrNbO$_{3.33}$  (n = 6) and insulating SrNbO$_{3.5}$ (n = 4) compounds. While the number of oxygen atoms and total electron density remains nominally (almost) the same as SrNbO$_{3.4}$, this alternate structure contains units of fully oxidated (n = 4) SrNbO$_{3.5}$ which is a ferroelectric with extraordinary high values of polarizability (at room temperature, $\varepsilon_{1a}=75, \varepsilon_{1b}=43, \varepsilon_{1c}=46$) \cite{polar}. In addition, our DFT calculations show that the ground state free energy for standard SrNbO$_{3.4}$ (n = 5) and intercalated SrNbO$_{3.4}$ (n = 4, 6, ...) configurations are very close (\SI{-506.36}{\electronvolt} and \SI{-506.28}{\electronvolt}, respectively), implying that this new configuration is stable. Moreover, this SrNbO$_{3.4}$ configuration may also be a better representative of the experimental oxygen-rich SrNbO$_{3+\delta}$ samples, which showed some signs of similar non-uniformity in its growth \cite{asmara}.

\begin{figure}
	\includegraphics[width=8.6 cm,height=8cm]{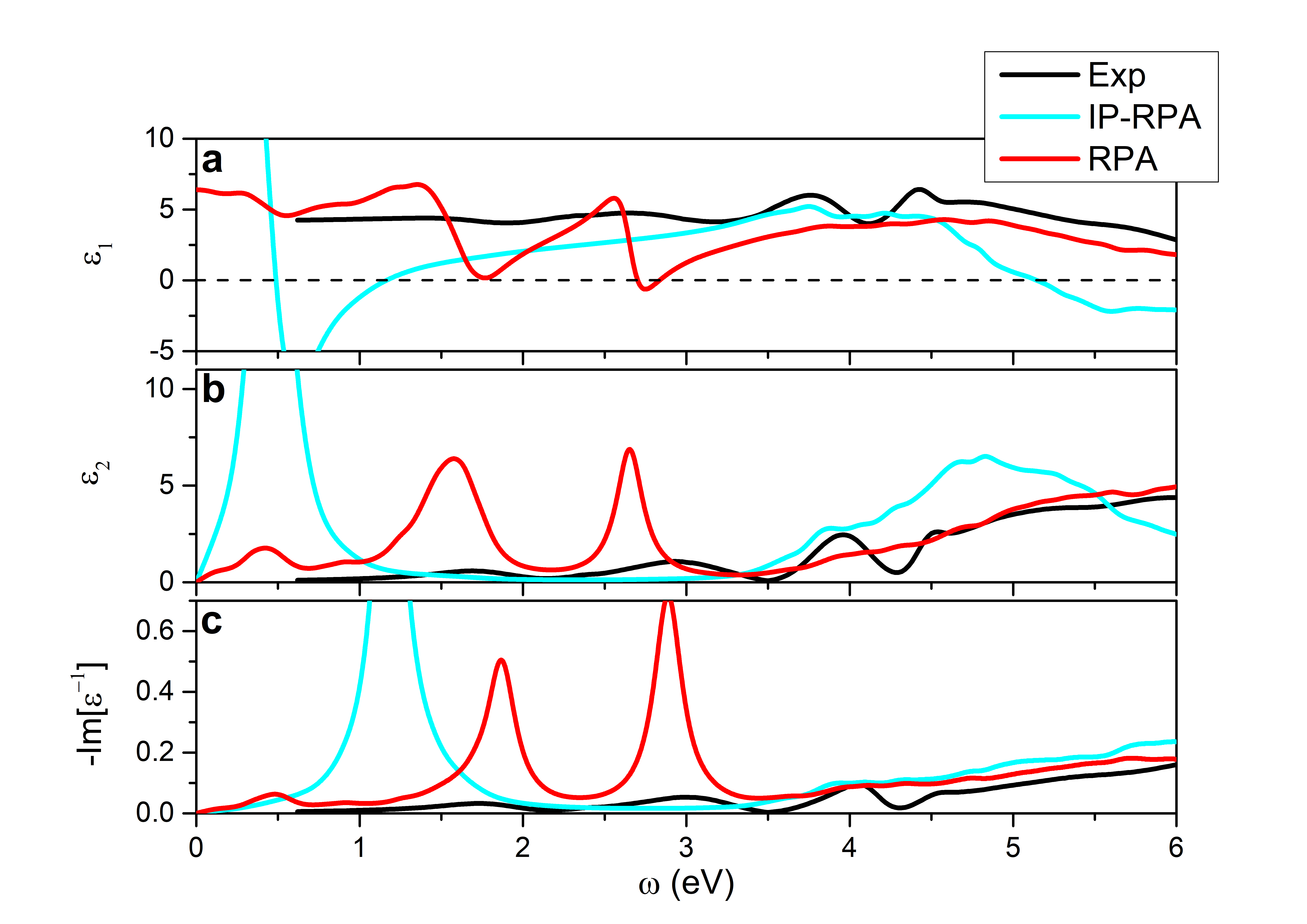}
	\caption{Calculated macroscopic complex dielectric and loss functions of SrNbO$_{3.4}$ with alternate structure along the c-axis. The calculated (a) real part, $\varepsilon_1$, and (b) imaginary part, $\varepsilon_2$, of complex dielectric function as well as (c) loss function, -Im$[\varepsilon^{-1}]$, of SrNbO$_{3.4}$ with alternate structure (n = 4, 6, ...) are compared with those of experiment (Exp) \cite{asmara}. RPA and IP-RPA refer to the calculation with and without LFEs, respectively. The horizontal dashed line indicates the zero of $\varepsilon_1$ to pinpoint the conventional plasmon peak energy.}
	\label{fig6}
\end{figure}

As Fig. \ref{fig6} shows, the IP-RPA loss function of SrNbO$_{3.4}$ (n = 4, 6, ...) is almost unchanged in comparison with the standard SrNbO$_{3.4}$ (n = 5) because they have almost the same charge density. In contrast, in the RPA + LFEs results, three main peaks appear in both dielectric and loss functions at almost the same energies of $\sim$ \SI{0.9}{\electronvolt}, $\sim$ \SI{2.0}{\electronvolt}, and $\sim$ \SI{3.3}{\electronvolt}, even more consistent with experimental data. For comparison, in the RPA + LFEs calculations using standard (n = 5) structure (Fig. \ref{fig3}), only two prominent loss function peaks appear within the experimental measurement limit of \SIrange{0.6}{6.0}{\electronvolt} and these peaks are $\sim$ \SI{0.3}{\electronvolt} blue-shifted compared to the corresponding dielectric function peaks in $\varepsilon_2$. The absence of these peak shifts in the alternate structure implies stronger LFEs, which suppress the long-range Coulomb interaction at the response function kernel in TDDFT \cite{sottile}. This can be expected from the highly polarizable (n = 4) SrNbO$_{3.5}$ units embedded within. The intensities of calculated peaks are also less intense compared to the standard structure, which allows these anomalous plasmons to have low loss, consistent with experimental data. This intensity damping is attributed to the repulsive nature of LFEs (as a Coulomb interaction) between charge density waves (CDW) \cite{lfe5}. This means that the anomalous plasmon peaks in SrNbO$_{3.4}$ do indeed arise from the LFEs.

\section{Discussion}

We have shown that the LFEs are essential to create multiple unconventional plasmons with spectral weight transfers. The calculated peaks do appear to be globally red-shifted compared with experimental data. This can be due to the fact that electron-electron and electron-hole interactions are neglected in RPA. Nevertheless, SrNbO$_{3.4}$ has metallic behavior which suppresses these interactions. From our calculations and in comparison to experimental data, it can be clearly inferred that the LFEs are the main cause of the spectral weight transfer from single conventional plasmon peak to higher energies with the subsequent creation of multiple plasmon peaks.
\begin{figure}[htbp]
	\centering
	\begin{minipage}{8 cm}
		\includegraphics[width=8.6 cm]{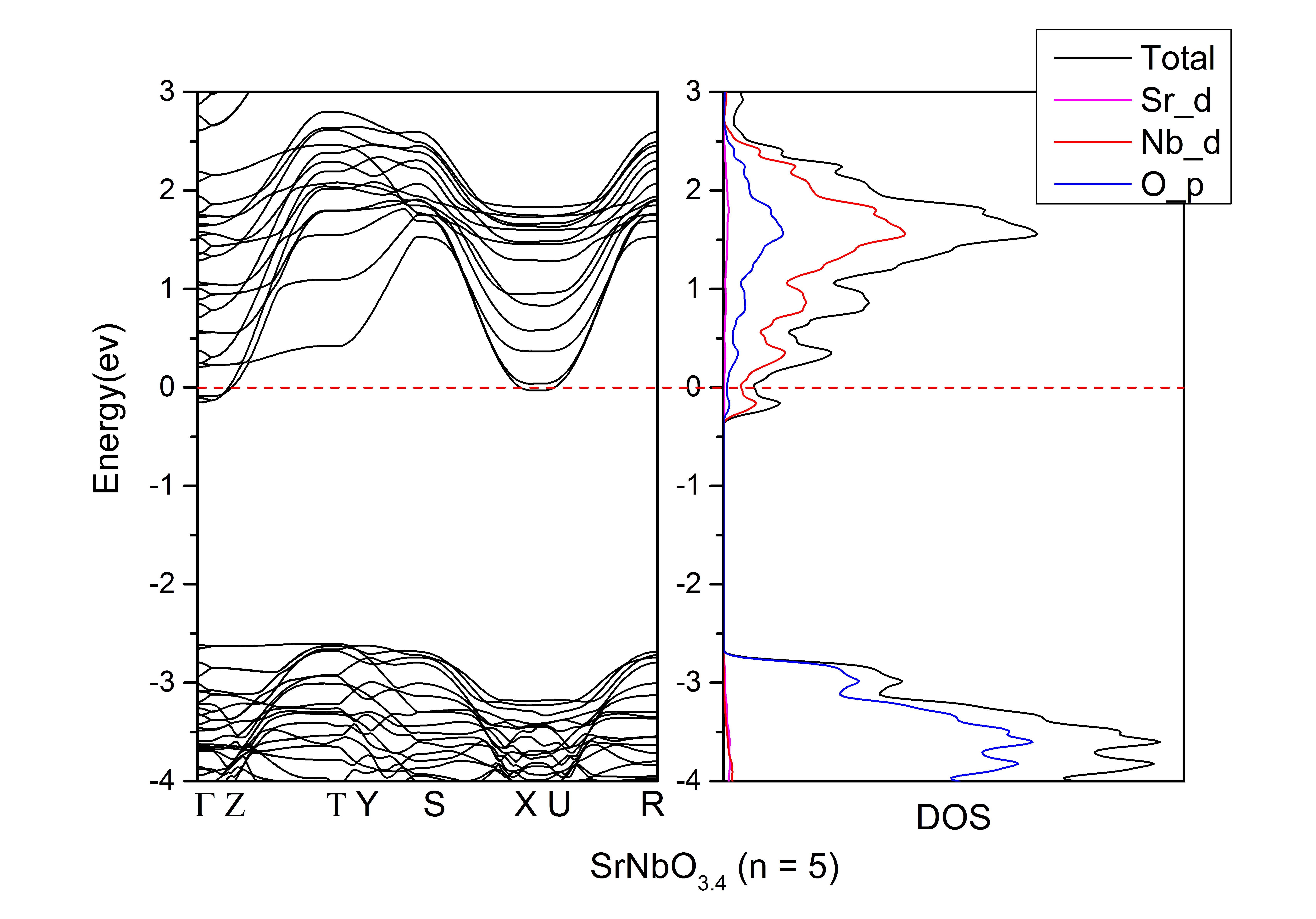}
	\end{minipage}
	\begin{minipage}{8 cm}
		\includegraphics[width=8.6 cm]{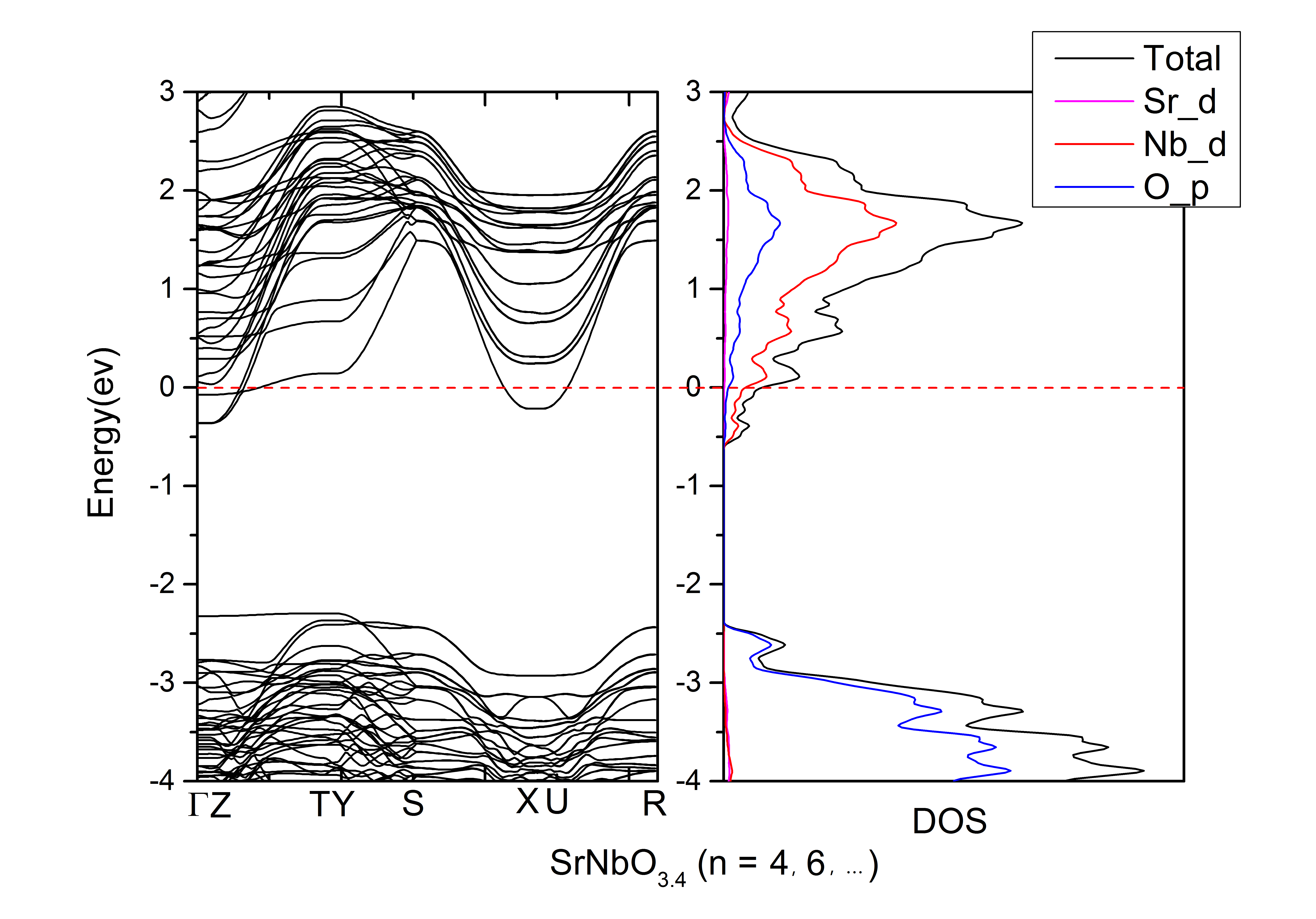}
	\end{minipage}
	\caption{Band structure and density of states of SrNbO$_{3.4}$ (n = 5) and SrNbO$_{3.4}$ (n = 4, 6, ...). Red dashed line represents the Fermi level.}
	\label{fig7}
\end{figure}
	
The next question is related to which bands contribute to the creation of these unconventional plasmons. In Fig. \ref{fig7}, the DFT band structures and DOS of the introduced two types of SrNbO$_{3.4}$ are shown. Similar to SrNbO$_3$, both configurations of SrNbO$_{3.4}$ show a metallic feature as the $E_F$ crosses the conduction band. For these two types oxygen-rich materials, the $E_F$ is placed at the bottom of the conduction band which indicates a lower conductivity than that in SrNbO$_3$. The direct band gap $\sim$ \SI{2.3}{\electronvolt} is found at the $\Gamma$ point and it is just below the $E_F$ in both configurations. From IP-RPA results, the plasmon peak is placed at $\sim$ \SI{1.2}{\electronvolt}, which implies that the peaks in the loss function are not likely originated from interband transitions alone. Moreover, it should be noted that the multiple plasmon frequencies as shown in the RPA spectra are created by a mixture of different independent transitions when the LFEs are taken into account. Thus, we infer that these multiple plasmons are still collective excitations but involving structural confinement of Nb 4$d$ electrons where $d-d$ transitions play a major role. On the other hand, a strong modification of the spectra from LFEs suggests a strong electronic inhomogeneity along the c-axis. Previous experimental and numerical reports \cite{asmara,kuntscher1,kuntscher2,polar} show that the density of occupied states at the $E_F$ mainly comes from the central Nb atoms located in the middle of the slabs whereas the contributions of the other two Nb sites decrease substantially towards the edges. This can be an indication that the EOP is able to confine the free Nb 4$d$ electrons to stay at the center area of slabs and generate strong LFEs.

To further investigate the effects of structural confinement caused by the EOP, we consider the anisotropic structure of oxygen-rich SrNbO$_{3+\delta}$. Unlike the cubic perovskite structure of SrNbO$_3$, the structure of SrNbO$_{3.4}$ is highly anisotropic. The EOP separates different layers along the c-axis which preclude the free electrons transport to the neighboring layers.  On the other hand, as can be seen in Fig. \ref{fig4}, within each slab the NbO$_6$ octahedra extend zig-zag-like along the b-axis and chain-like along the a-axis. This strong structural anisotropy makes SrNbO$_{3.4}$ a low dimensional metal \cite{kuntscher1,kuntscher2}. From the band structure as shown in Fig. \ref{fig7}, the $E_F$ crosses the conduction band along $\Gamma - Z$ and $X-U$ directions. At the same time, there is no apparent band dispersion along these two directions. This indicates an insulating feature along the c-axis, agree with the experimental results that the conductivity along the c-axis is quite low \cite{kuntscher1,lichtenberg}.
\begin{figure}[htbp]
	\includegraphics[width=8.6 cm]{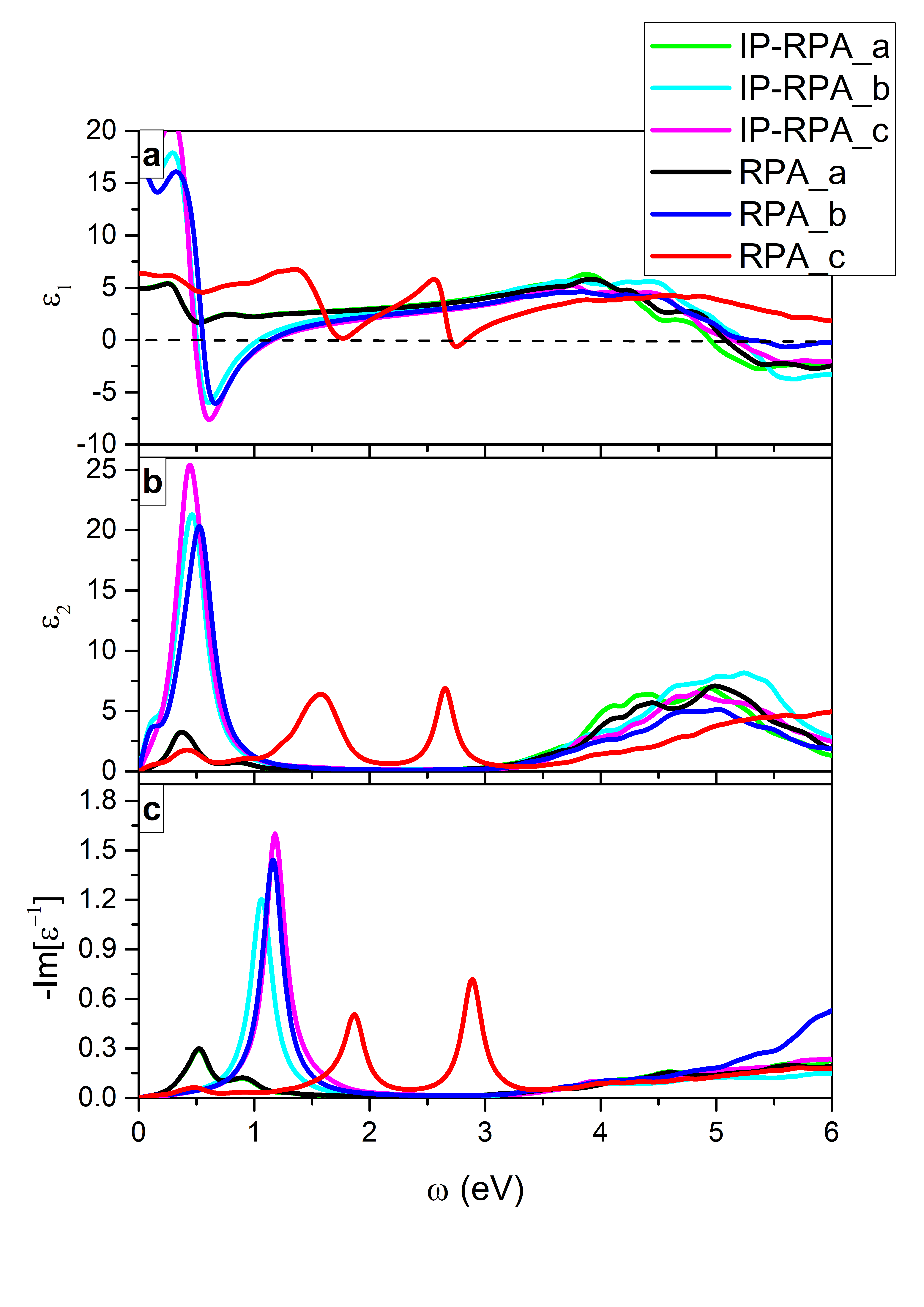}
	\caption{Calculated macroscopic complex dielectric and loss functions of SrNbO$_{3.4}$ ($n=5$) along the a-, b-, and c-axis. (a) real part, $\varepsilon_1$, (b) imaginary part, $\varepsilon_2$, of complex dielectric function and (c) loss function, -Im$[\varepsilon^{-1}]$. RPA and IP-RPA refer to the calculation with and without LFEs, respectively. The horizontal dashed line indicates the zero of $\varepsilon_1$ to pinpoint the conventional plasmon peak energy.}
	\label{fig8}
\end{figure}

In Fig. (\ref{fig8}) we show the calculated dielectric and loss function of SrNbO$_{3.4}$ (n = 5) along different axes. It can be seen that for the electric field along the $a$- and $b$-axis, the IP-RPA and RPA calculations almost give an identical spectrum. This implies that LFEs have little impact on spectra along these directions and thus further highlights the importance of the EOP in creating strong LFEs. Interestingly, although the $a$-axis keeps the same Nb-O chain-like structure as isotropic SrNbO$_3$, the $\varepsilon_1$ stays positive which indicates an insulating character. This is consistent with the previous report \cite{kuntscher1} that a unique small energy gap at the Fermi level exists in this conducting chain at low temperature. Further analyses show that this temperature-driven metal to insulator transition may be created by the formation of a CDW below the Peierls transition \cite{kuntscher1, kuntscher2}. Under IP-RPA, the b-axis and the c-axis have similar spectra. A conventional type plasmon is shown in both axes at $\sim$ \SI{1.2}{\electronvolt}. However, when LFEs are included, only the plasmon peak along the c-axis transfers to higher energy and splits into two branches located at $\sim$ \SI{1.9}{\electronvolt} and $\sim$ \SI{2.9}{\electronvolt}, while the loss function along the b-axis only has a slight blue-shift. This indicates that the confinement caused by EOP is crucial in creating unconventional multiple plasmons. As an interesting comparison, a strong modification of the loss function caused by LFEs has been found in rutile TiO$_2$ where semi-core Ti 3$p$ electrons are involved \cite{vast}. However, as we have shown in Fig. \ref{fig3}, opposite to the SrNbO$_{3+\delta}$ case, the partially filled Nb 4$d$ orbitals do not induce strong LFEs in SrNbO$_3$. Instead, the strong localization of Nb 4$d$ electrons in oxygen-rich SrNbO$_{3+\delta}$ samples are induced by the EOP which play a pivotal role in creating multiple unconventional plasmons in the experiment.

\begin{figure}[htbp]
	\centering
	\includegraphics[width=8.6 cm]{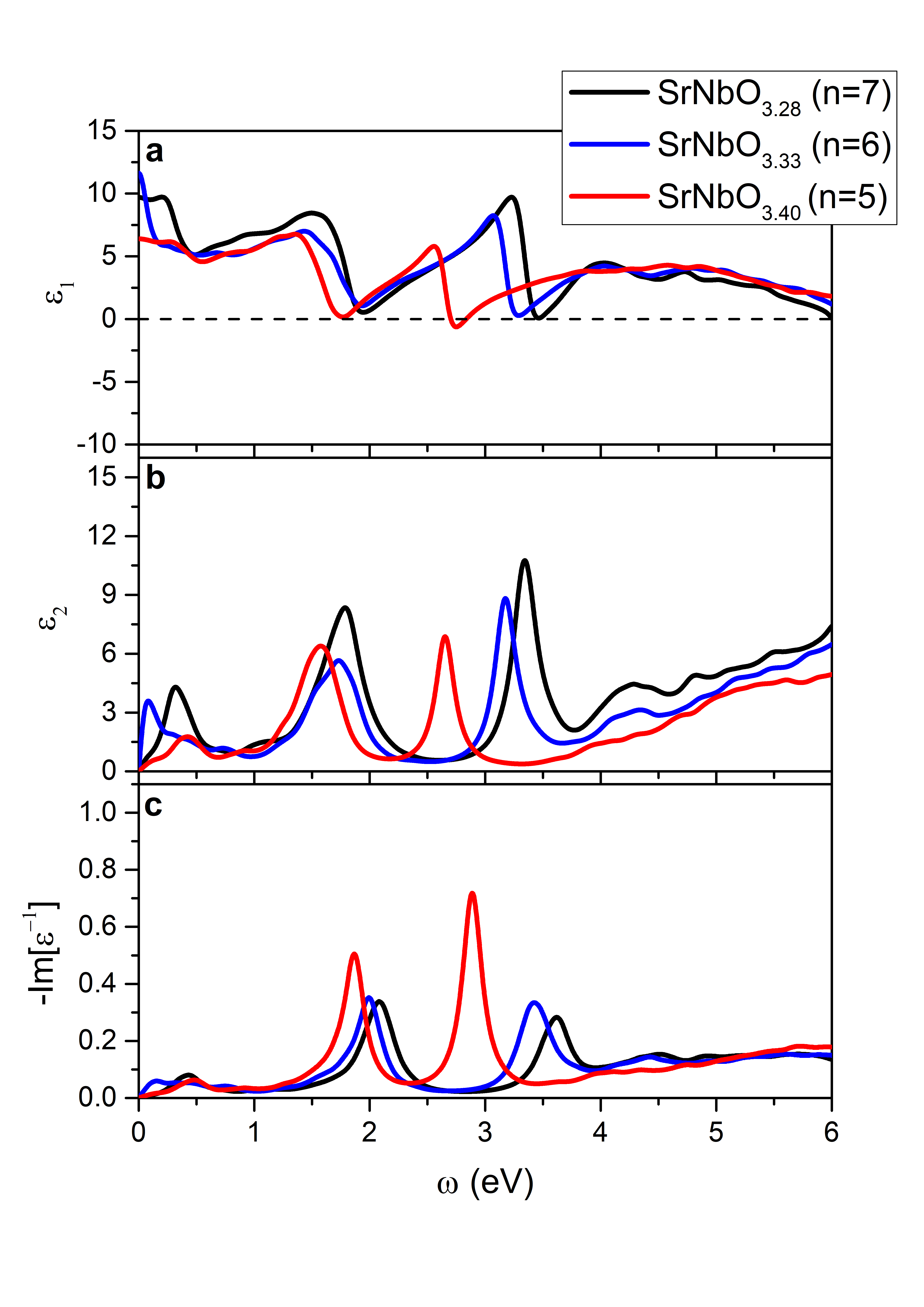}
	\caption{Calculated macroscopic RPA dielectric and loss functions of SrNbO$_{3+\delta}$ with different oxygen content. (a) real part, $\epsilon_1$, (b) imaginary part, $\epsilon_2$, of complex dielectric function and (c) loss function, -Im$[\varepsilon^{-1}]$. The horizontal dashed line indicates the zero of $\varepsilon_1$ to pinpoint the conventional plasmon peak energy.}
	\label{fig9}
\end{figure}

Another influence from EOP can be obtained by comparing results from SrNbO$_{3+\delta}$ with different oxygen content. The structure of oxygen-rich SrNbO$_{3+\delta}$ compounds is strongly affected by oxygen content and the thickness of each layer is given by the number of NbO$_6$ octahedra. As shown in Fig. \ref{fig1}, the thickness of each layer, which is separated by EOP, can be tuned by varying oxygen content. Without EOP, an infinity member of Sr$_n$Nb$_n$O$_{3n+2}$ realizes SrNbO$_3$. Figure \ref{fig9} shows the calculated RPA dielectric and loss functions of the layered strontium niobates with different oxygen content. It can be seen that the intensity and frequencies of unconventional plasmons are affected by oxygen content, which further reflects the influence of the distance between two neighboring EOPs. Both dielectric and loss function can be altered by the thickness of slabs. The peaks of $\varepsilon_2$ and loss function have negative correlations with oxygen content. This is consistent with the fact that the increase of oxygen content reduces the free charge density. One could expect a similar spectra shift can be achieved by introducing strain to reduce or enlarge the distance between neighboring EOP. The $\varepsilon_1$ of SrNbO$_{3.4}$ ($n=5$) has negative values at the plasmon energies $\sim$ \SI{1.9}{\electronvolt} and $\sim$ \SI{2.9}{\electronvolt}, which show a conventional character. However, the real dielectric function $\varepsilon_1$ of SrNbO$_{3.28}$ ($n=7$) and SrNbO$_{3.33}$ ($n=6$) (as well as SrNbO$_{3.4}$ ($n=4, 6, ...$)) stays positive from \SIrange{0.6}{6.0}{\electronvolt}. In this case, these plasmon energies do not coincide with the zero but instead coincide with the local minima of $\varepsilon_1$.

\section{Summary}
We have analyzed the complex dielectric and loss functions of strontium niobate compounds of SrNbO$_{3}$ and SrNbO$_{3.4}$ and the alternating superstructure of SrNbO$_{3.3}$ and SrNbO$_{3.5}$. By performing fully \textit{ab initio} calculations, we show that most of the unconventional plasmons behaviors in SrNbO$_{3.4}$ can be reproduced within RPA when local field effects (LFEs) are considered. In the isotropic metallic SrNbO$_{3}$, a conventional plasmon is present and determined by both interband and intraband transitions. In insulating-like SrNbO$_{3.4}$, the conventional plasmon is suppressed by LFEs and through spectral weight transfer multiple frequencies and low loss plasmons are generated. Consistent with experimental data, these plasmons are determined by an interplay of charge inhomogeneity induced by oxygen doping and high-polarizability induced by non-uniformity growth of sample. From our DFT analysis, we have inferred that these unconventional plasmons are collective excitations of electrons confined by the EOP. A strong anisotropy of the dielectric and loss functions is found. We further show that the frequencies of these plasmons can be tuned. Our result shows that this plasmon phenomena with multiple resonances can be extended to other perovskite materials with strong electrical inhomogeneity.

\section{acknowledgments}

This work is supported by the Ministry of Educations (MOE2015-T2-2-065, MOE2015-T2-1-099, and MOE2015-T2-2-147), Singapore National Research Foundation under its Competitive Research Funding (NRF-CRP 8-2011-06, R-398-000-087-281 and NRF-CRP15-2015-01) and under its Medium Sized Centre Programme (Centre for Advanced 2D Materials and Graphene Research Centre), NUS YIA, and FRC (R-144-000-379-114 and R-144-000-368-112). T.Z. and P.E.T. contribute equally to this work.


\begin{thebibliography}{[Vo]}
   	\bibitem{ridgley}
   D. Ridgley and R. Ward, J. Am. Chem. Soc. \textbf{77}, 6132-6136 (1955).
   \bibitem{isawa}
    K. Isawa, J. Sugiyama, K. Matsuura, A. Nozaki, and H. Yamauchi, Phys. Rev. B \textbf{47}, 2849-2853 (1993).	
   \bibitem{weber}
   J. E. Weber, C. Kegler, N. B\"uttgen, H. A. Krug von Nidda, A. Loidl, and F. Lichtenberg, Phys. Rev. B \textbf{64}, 235414 (2001).
   \bibitem{sakai}
   A. Sakai, T. Kanno, K. Takahashi, Y. Yamada, and H. Adachi, J. Appl. Phys. \textbf{108}, 103706 (2010).
   \bibitem{kobayashi}
   W. Kobayashi, Y. Hayashi, M. Matsushita, Y. Yamamoto, I. Terasaki, A. Nakao, H. Nakao, Y. Murakami, Y. Moritomo, H. Yamauchi, and M. Karppinen, Phys. Rev. B \textbf{84}, 085118 (2011).
   \bibitem{lichtenberg}
   F. Lichtenberg, A. Herrnberger, K. Wiedenmann, and J. Mannhart, Prog. Solid State Chem. \textbf{29}, 1-70 (2001).
   \bibitem{kuntscher1}
   C. A. Kuntscher, S. Schuppler, P. Haas, B. Gorshunov, M. Dressel, M. Grioni, F. Lichtenberg, A. Herrnberger, F. Mayr, and J. Mannhart, Phys. Rev. Lett. \textbf{89}, 236403 (2002).
   \bibitem{kuntscher2}
   C. A. Kuntscher, S. Schuppler, P. Haas, B. Gorshunov, M. Dressel, M. Grioni, and F. Lichtenberg, Phys. Rev. B \textbf{70}, 245123 (2004).
   \bibitem{chen}
   C. Chen, S. Lv, Z. Wang, K. Akagi, F. Lichtenberg, Y. Ikuhara, and J. G. Bednorz, Appl. Phys. Lett. \textbf{105}, 221602 (2014).
   \bibitem{xu}
   X. Xu, C. Randorn, P. Efstathiou, and J. T. S. Irvine, Nat. Mater. \textbf{11}, 595-598 (2012).
   \bibitem{wan}
   D. Y. Wan, Y. L. Zhao, Y. Cai, T. C. Asmara, Z. Huang, J. Q. Chen, J. Hong, S. M. Yin, C. T. Nelson, M. R. Motapothula, B. X. Yan, D. Xiang, X. Chi, H. Zheng, W. Chen, R. Xu, Ariando, A. Rusydi, A. M. Minor, M. B. H. Breese, M. Sherburne, M. Asta, Q. H. Xu, and T. Venkatesan, Nat. Commun. \textbf{8}, 15070 (2017).
  \bibitem{asmara}
   T. C. Asmara, D. Y. Wan, Y. L. Zhao, M. A. Majidi, C. T. Nelson, M. C. Scott, Y. Cai, B. X. Yan, D. Schmidt, M. Yang, T. Zhu, P. E. Trevisanutto, M. R. Motapothula, Y. P. Feng, M. B. H. Breese, M. Sherburne, M. Asta, A. Minor, T. Venkatesan, and A. Rusydi, Nat. Commun. \textbf{8}, 15271 (2017).
  \bibitem{dmft1}
   A. Georges, G. Kotliar, W. Krauth, and M. J. Rozenberg, Rev. Mod. Phys. \textbf{68}, 13 (1996).
  \bibitem{dmft2}
   G. Kotliar, S. Y. Savrasov, K. Haule, V. S. Oudovenko, O. Parcollet, and C. A. Marianetti, Rev. Mod. Phys. \textbf{78}, 865-951 (2006).
   \bibitem{vanloon}
   E. G. C. P. Van Loon, H. Hafermann, A. I. Lichtenstein, A. N. Rubtsov, and M. I. Katsnelson, Phys. Rev. Lett. \textbf{113}, 246407 (2014).
   \bibitem{adler}
   S. L. Adler, Phys. Rev. \textbf{126}, 413 (1962).
   \bibitem{wiser}
   N. Wiser, Phys. Rev. \textbf{129}, 62 (1963).
  \bibitem{lfe1}
   S. G. Louie, J. R. Chelikowsky, and M. L. Cohen, Phys. Rev. Lett. \textbf{34}, 155-158 (1975).
  \bibitem{lfe2}
  K. Sturm, Phys. Rev. Lett. \textbf{40}, 1599-1602 (1978).
  \bibitem{lfe3}
  F. Aryasetiawan, O. Gunnarsson, M. Knupfer, and J. Fink, Phys. Rev. B \textbf{50}, 7311 (1994).
  \bibitem{lfe4}
  S. Waidmann, M. Knupfer, B. Arnold, J. Fink, A. Fleszar, and W. Hanke, Phys. Rev. B \textbf{61}, 10149 (2000).
  \bibitem{lfe5}
  P. Cudazzo, M. Gatti, and A. Rubio, New J. Phys. \textbf{15}, 125005 (2013).
  \bibitem{vast}
  N. Vast, L. Reining, V. Olevano, P. Schattschneider, and B. Jouffrey, Phys. Rev. Lett. \textbf{88}, 037601 (2002).	
  \bibitem{gajdos}
  M. Gajdo\v s, K. Hummer, G. Kresse, J. Furthmüller, and F. Bechstedt, Phys. Rev. B \textbf{73}, 045112 (2006).
  \bibitem{rpa}
  H. Ehrenreich, M. H. Cohen, Phys. Rev. \textbf{115}, 786 (1959).
  \bibitem{Vasp1}
   G. Kresse and J. Furthmüller, Phys. Rev. B \textbf{54}, 11169 (1996).
  \bibitem{Vasp2}
   G. Kresse and D. Joubert, Phys. Rev. B \textbf{59}, 1758 (1999).
  \bibitem{paw}
  P. E. Bl\"ochl, Phys. Rev. B, \textbf{50}, 17953, (1994).
  \bibitem{gga}
  J. P. Perdew, K. Burke, and M. Ernzerhof, Phys. Rev. Lett., \textbf{77}, 3865, (1996).
  \bibitem{ks}
  W. Kohn, and L. Sham. Phys. Rev. \textbf{140}, A1133 (1965)
  \bibitem{mp}
  H. J. Monkhorst, and J. D. Pack, Phys. Rev. B \textbf{13}, 5188 (1976).
  \bibitem{Onida}
  G. Onida, L. Reining, and A. Rubio, Rev. Mod. Phys. \textbf{74}, 601-659 (2002).
  \bibitem{sottile}
  F. Sottile, F. Bruneval, A. G. Marinopoulos, L. K. Dash, S. Botti, V. Olevano, N. Vast, A. Rubio, and L, Reining, Int. J. Quantum Chem. \textbf{102}, 684-701 (2005).
  \bibitem{dp}
  M. Cazzaniga, L. Caramella, N. Manini, and G. Onida, Phys. Rev. B \textbf{82} 035104 (2010), www.dp-code.org
  \bibitem{gatti}
  M. Gatti, F. Sottile, and L. Reining, Phys. Rev. B \textbf{91} 195137 (2015)
  \bibitem{polar}
  C. A. Kuntscher, S. Gerhold, N. N\"ucker, T. R. Cummins, D. H. Lu, S. Schuppler, C. S. Gopinath, F. Lichtenberg, J. Mannhart, and K. P. Bohnen, Phys. Rev. B \textbf{61} 1876 (2000)
\end{thebibliography}
\end{document}